\documentclass[aps,nofootinbib,amsmath,prc,showpacs,groupedaddress,10pt,superscriptaddress,notitlepage]{revtex4-1}

\usepackage[paperwidth=210mm,paperheight=297mm,centering,hmargin=2.4cm,tmargin=2.8cm,bmargin=3.5cm]{geometry}

\usepackage{amsfonts}
\usepackage{amsmath}
\usepackage{amssymb}
\usepackage{empheq}
\usepackage{epsfig}
\usepackage{latexsym}
\usepackage{paralist}
\usepackage{fancyhdr}
\usepackage{graphicx}

\numberwithin{equation}{section}

\usepackage[vcentermath]{youngtab}
\usepackage{young}
\usepackage{ytableau}
\usepackage{etex}
\usepackage{braket}
\usepackage{float}
\usepackage{slashed}
\usepackage{xcolor}
\usepackage{subcaption}
\usepackage{soul}

\usepackage{dcolumn}

\makeatletter
\@addtoreset{equation}{section}

\makeatother

\def\bea{\begin{eqnarray}} 
\def\eea{\end{eqnarray}}
\def\be{\begin{equation}} 
\def\ee{\end{equation}} 
\def\ba{\begin{array}}
\def\ea{\end{array}} 
\def\nn{\nonumber}

\def\be{\begin{equation}}
\def\ee{\end{equation}}
\def\bea{\begin{eqnarray}}
\def\eea{\end{eqnarray}}

\def\nn{\nonumber}
\def\ep{\epsilon}

\def\cS{{\cal S}}

\renewcommand{\thefootnote}{\fnsymbol{footnote}}

\let\oldtitle\title
\renewcommand{\title}[1]{\oldtitle{\color{blue}{#1}}}

\let\oldeqref\eqref
\let\oldcite\cite
\renewcommand{\eqref}[1]{{\color{blue}\oldeqref{#1}}}
\renewcommand{\cite}[1]{{\color{blue}\oldcite{#1}}}

\makeatletter
\let\reftagform@=\tagform@
\def\tagform@#1{\maketag@@@{\ignorespaces\textcolor{blue}{(\ignorespaces #1 \unskip\@@italiccorr \ignorespaces)\ignorespaces}}}
\makeatother

\makeatletter
\renewcommand{\p@subsection}{}
\renewcommand{\p@subsubsection}{}
\makeatother

\usepackage{mathpazo}
\usepackage{amsmath}

\begin{document}

\title{Multi-critical multi-field models: a CFT approach to the leading order.}

\author{A.\ Codello}
\email{a.codello@gmail.com}
\affiliation{Department of Physics, Southern University of Science and Technology, Shenzhen 518055, China}

\author{M.\ Safari}
\email{safari@bo.infn.it}
\affiliation{INFN - Sezione di Bologna, via Irnerio 46, 40126 Bologna, Italy}
\affiliation{
Dipartimento di Fisica e Astronomia,
via Irnerio 46, 40126 Bologna, Italy}

\author{G.\ P.\ Vacca}
\email{vacca@bo.infn.it}
\affiliation{INFN - Sezione di Bologna, via Irnerio 46, 40126 Bologna, Italy}

\author{O.\ Zanusso}
\email{omar.zanusso@uni-jena.de}
\affiliation{
Theoretisch-Physikalisches Institut, Friedrich-Schiller-Universit\"{a}t Jena,
Max-Wien-Platz 1, 07743 Jena, Germany}
\affiliation{INFN - Sezione di Bologna, via Irnerio 46, 40126 Bologna, Italy}

\begin{abstract}
We present some general results for the multi-critical multi-field models in $d\!>\!2$ recently obtained using CFT and Schwinger-Dyson methods at perturbative level without assuming any symmetry. Results in the leading non trivial order are derived consistently for several conformal data in full agreement with functional perturbative RG methods. Mechanisms like emergent (possibly approximate) symmetries can be naturally investigated in this framework.
\end{abstract}

\maketitle

\renewcommand{\thefootnote}{\arabic{footnote}}
\setcounter{footnote}{0}

 \section{Introduction}
Quantum and Statistical Field Theories are important mathematical models which can be used to describe physical systems and their universal behavior 
approaching criticality. The theoretical paradigm was strongly developed in the last decades starting from modern Renormalization Group (RG) 
concepts~\cite{Wilson:1971bg, Wilson:1971dh, Wilson:1973jj}, under which a critical theory is seen as a scale invariant fixed point of the RG flow~\cite{Cardy:1996xt}. 
Another line of interesting developments followed the observation and (partial) understanding that Poincar\'e and scale invariance for unitary theories can be lifted 
to the larger conformal symmetry. This seems to work not only in four dimension since there is a strong evidence to be true also in three dimension (where 
Conformal Bootstrap methods~\cite{Rattazzi:2008pe,ElShowk:2012ht} give numerical predictions for the critical exponents with unmatched precision). This 
enlargement of symmetry seems to be true, at least up to the present investigations, also for non unitary theories and in other dimensionalities, including fractional 
ones. Indeed the assumption of conformal symmetry leads to correct results at least within the approximations adopted. Many investigations  in a Conformal Field 
Theory (CFT) framework have been carried on with perturbative methods, e.g. \ in the $\ep$-expansion, originally developed in RG analysis~\cite{Wilson:1971dc},
taking advantage of the knowledge of the equation of motion at criticality~\cite{Rychkov:2015naa, Basu:2015gpa, Nakayama:2016cim, Nii:2016lpa, 
Hasegawa:2016piv, Codello:2017qek, Codello:2018nbe, Antipin:2019vdg}, using the Mellin space approach~\cite{Gopakumar:2016wkt}, 
large N~\cite{Rong:2017cow} or large spin~\cite{Alday:2016njk} expansions, conformal block expansion~\cite{Gliozzi:2016ysv}, etc. 
Expecially RG methods have also revealed as a fertile source for applications to other quantitative sciences.
Moreover these methods are useful not only to investigate the effective behavior of different physical systems at large distances, but are important in the quest of 
defining which quantum theories can be considered consistent and fundamental, a question still open in particle physics.

In what follows we shall illustrate how to conveniently use few basic properties of CFTs in $d>2$, namely its constraints on the two and three point correlators, following~\cite{Codello:2017qek, Safari:2017irw, Codello:2018nbe}.
Adopting a basis $O_a$ of normalized scalar primary operators~\footnote{An operator is said primary when, taken in the origin, commutes with, or "is annihilated by", the special conformal generator.}  with scaling dimensions $\Delta_a$,
the two-point correlators have the form
\begin{align}\label{cft-2pf}
\braket{O_a(x)  O_b(y)} =\frac{\delta_{ab}} {|x-y|^{2 \Delta_a}}
\end{align}
and the three-point correlator for scalar primary operators is also strongly constrained by conformal symmetry and reads
\begin{align}\label{cft-3pf}
\braket{O_a(x)  O_b(y) O_c(z)} =\frac{C_{abc} }{|x-y|^{\Delta_{a}+\Delta_{b}-\Delta_{c}} 
|y-z|^{\Delta_{b}+\Delta_{c}-\Delta_{a}} |z-x|^{\Delta_{c}+\Delta_{a}-\Delta_{b}}} \,,
\end{align}
in which $C_{abc}$ are the structure constants of the CFT. The quantities $\{\Delta_a, C_{abc} \}$ are also known as conformal data and
the CFT is completely determined by their knowledge.

Given the set of fields (and eventually symmetries) characterizing the Quantum Field Theory QFT (or the CFT), perturbation theory is a powerful tool to get
an idea of the critical points in the theory space. In particular the perturbative $\ep$-expansion analysis below the upper critical dimension,
at which the theory is trivial, is very effective. In this case another useful step, helping in simplifying the extraction of the first non trivial corrections
to conformal data, is the adoption of a Lagrangian description ($S\!=\!\int \!d^d x {\cal L}$) at criticality which allows to use the equation of motion
through the Schwinger-Dyson Equations (SDE)  for correlators. Since at separate points no contact terms are present, one has
\begin{align}
\left\langle\frac{\delta S}{\delta\phi}(x) \, O_1(y) \, O_2 (z) \dots\right\rangle  = 0.
\label{prop}
\end{align}
It is interesting to consider this CFT-SDE approach in the study of multi-critical theories of $N$ fields characterized by a critical generic potential
\be
V=\frac{1}{m!} V_{i_1 \cdots i_m}\, \phi_{i_1} \cdots \phi_{i_m},
\label{mcpotential}
\ee
with $\binom{m+N-1}{m}$ independent monomial interactions (and couplings).
The critical dimension is $d_c=2m/(m-2)$, a fractional number for $m=5$ or $m>6$.
For theories with the simple standard kinetic term the critical action reads
\be
S=\int \! d^dx \left[{\textstyle{\frac{1}{2}}}\,\partial \phi_i \cdot \partial \phi_i + V(\phi) \right] 
\label{general_model}
\ee
and then all the fields have the same canonical dimension $\delta=d/2-1=2/(m-2)-\epsilon/2$ but may start to differ in the anomalous dimension $\gamma_a=\Delta_a-\delta$. Introducing $n=m/2$ one has $\delta=1/(n-1)-\epsilon/2$.

\begin{figure}[h]
\begin{center}
\includegraphics[width=4cm]{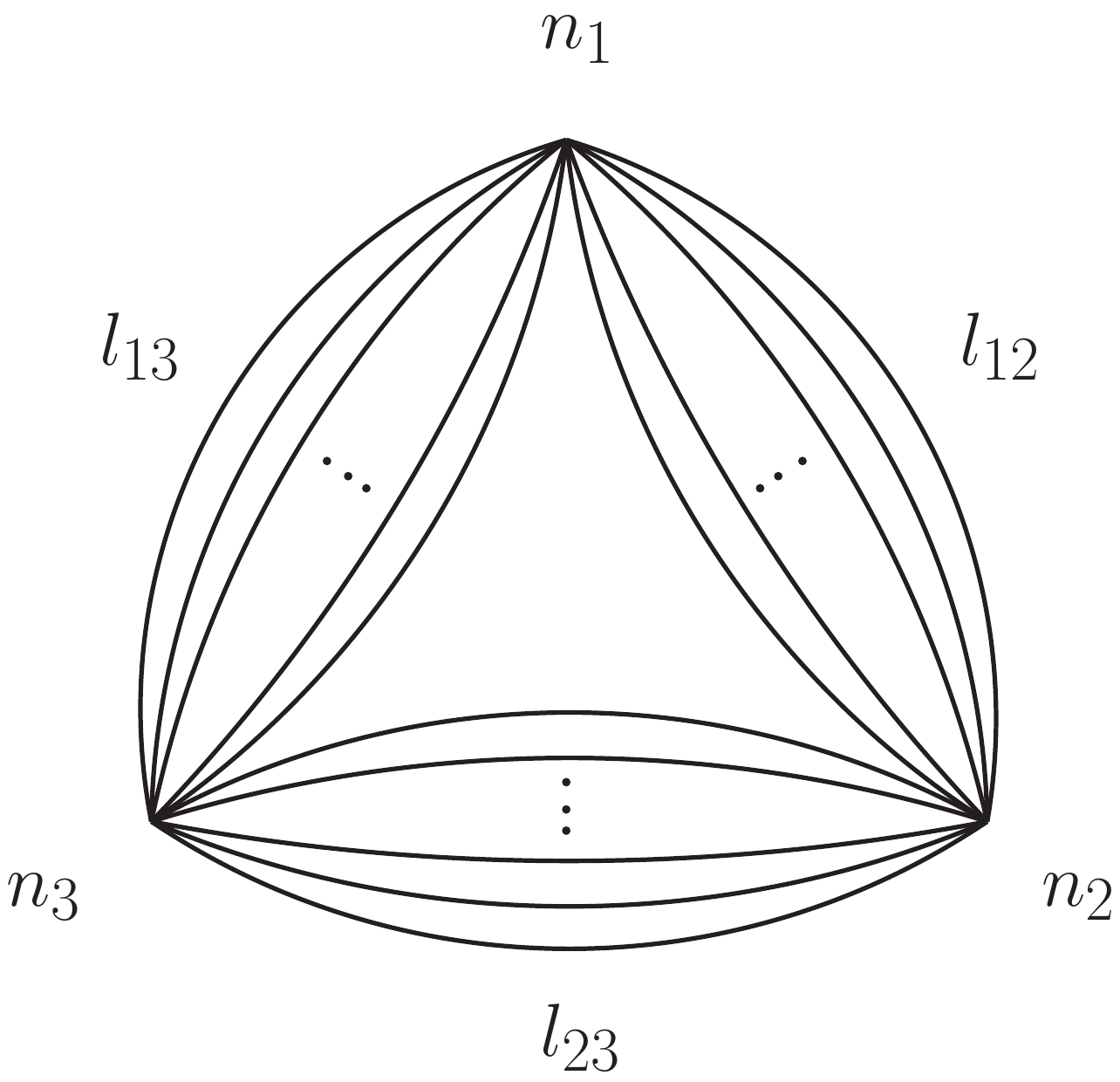}
\end{center}
\caption{Wick contraction counting of a three point correlator.
The vertices are labelled by $i=1,2,3$, the order of the $i$-th vertex is $n_i$, and there are $l_{ij}$ lines connecting two distinct vertices $i$ and $j$.}
\label{3pfcounting}
\end{figure}

The other ingredient needed for the perturbative analysis is the knowledge of the free theory correlators, which are defined at the upper critical dimension $d_c$ for $\ep=0$.
The two-point function is simply given by
\be \label{phiphi-free}
\braket{\phi_i(x)  \phi_j(y)} \overset{\mathrm{free}}=\frac{c\, \delta_{ij}}{|x-y|^{2 \delta_c}}\,,
\ee
where  $\delta_c=\frac{1}{2}d_c-1=1/(n\!-\!1)$ is the dimension of the field $\phi_i$ in the free theory ($\ep=0$) and
\be
c=\frac{1}{4 \pi} \frac{\Gamma(\delta_c)}{\pi^{\delta_c}}.
\label{c}
\ee
The two-point function can be put in the canonical form with a normalized coefficient using the rescaled field $\tilde{\phi}_i$ defined through $\phi_i=\sqrt{c}\tilde\phi_i$. Then the two-point function of the composite operators can be computed, e.g. 
\be
\braket{[\phi^{i_1}\cdots \phi^{i_k}](x)\,  [\phi_{j_1}\cdots \phi_{j_k}](y)} \overset{\mathrm{free}}{=}\delta^{i_1}_{(j_1}\cdots \delta^{i_k}_{j_k)} \,   \frac{k!\,c^k}{|x-y|^{2 k \delta_c}}\,,
\label{2pf-free}
\ee
in which on the r.h.s the $j$ indices are symmetrized (including the inverse factor of $k!$). 
We shall also need the expression for the generic three-point function 
\bea
&& \braket{[\phi_{i_1}\cdots \phi_{i_{n_1}}](x_1)\; [\phi_{j_1}\cdots \phi_{j_{n_2}}](x_2)\;[\phi_{k_1}\cdots \phi_{k_{n_3}}](x_3)}  \nn\\
&\overset{\mathrm{free}}{=}& 
  \frac{C_{i_1\cdots i_{n_1},j_1\cdots j_{n_2},k_1\cdots k_{n_3}}^{\mathrm{free}}}{|x_1\!-\!x_2|^{\delta_c (n_1+n_2-n_3)}|x_2\!-\!x_3|^{\delta_c (n_2+n_3-n_1)}
  |x_3\!-\!x_1|^{\delta_c (n_3+n_1-n_2)}} \,,
\label{3pf_free}
\eea
where the coefficients on the r.h.s are nonvanishing only when the number of propagators 
\be 
l_{ij} = \frac{1}{2}(n_i+n_j-n_k), \qquad i\neq j \neq k
\ee
in each edge of the diagram in Figure~\ref{3pfcounting} turns out to be nonnegative. 

They are obtained from the condition $n_i=l_{ij}+l_{ki}$ for $i\neq j \neq k$. In this case the coefficients are
\be 
C^{\mathrm{free}}_{i_1\cdots i_{n_1},j_1\cdots j_{n_2},k_1\cdots k_{n_3}} = C_{n_1,n_2,n_3}^{\mathrm{free}} \, (\delta_{i_1j_1} \cdots \delta_{i_{l_{12}}j_{l_{12}}}\,\delta_{i_{l_{12}+1}k_1} \cdots \delta_{i_{n_1}k_{l_{13}}}\,\delta_{k_{l_{13}+1}j_{l_{12}+1}} \cdots \delta_{k_{n_3}j_{n_2}}),
\label{3pfree}
\ee
where the parenthesis enclosing the Kronecker deltas indicates that the $i$s the $j$s and the $k$s are separately symmetrized (including an inverse factor of $l_{12}!l_{13}!l_{23}!$) and the first factor is just the single-field counterpart 
\be 
C_{n_1,n_2,n_3}^{\mathrm{free}} = 
 \frac{n_1! \ n_2! \ n_3!}{\left(\frac{n_1+n_2-n_3}{2}\right)!\left(\frac{n_2+n_3-n_1}{2}\right)!\left(\frac{n_3+n_1-n_2}{2}\right)!} c^{\frac{n_1+n_2+n_3}{2}} \,.
\label{c3_free}
\ee

We shall show that the systematic use of all the relations recalled above can give access to a large set of conformal data in the first non trivial order in the perturbative $\ep$-expansion~\cite{Codello:2017qek, Safari:2017irw, Codello:2018nbe}. The results are the same as those obtained with perturbative RG methods,
which if treated at functional level give rise to a very compact and effective computational framework. Actually for certain models (unitary multi-critical) one can easily reconstruct the Functional Perturbative RG (FPRG) equations~\cite{ODwyer:2007brp, Codello:2017hhh, Codello:2017epp, Safari:2017tgs, Zinati:2019gct} starting from the obtained CFT relations.

\noindent Before discussing which relations are implied by assuming the CFTs and the lagrangian description in the general multi-field case, let us illustrate how to deal with the simpler theories with a single scalar field~\cite{Codello:2017qek}. We shall restrict here to theories with standard kinetic term,
but investigations have been carried out also for multi-critical higher derivative theories revealing an unexpected rich structure~\cite{Safari:2017irw}.

\section{Review of the single scalar field case}
In this Section the field anomalous is denoted by $\gamma$ ($\Delta=\delta+\gamma$) while for the composite operators $[\phi^i]$ the scaling dimension is denoted by $\Delta_i=i\,\delta +\gamma_i$ (anomalous dimension $\gamma_i$). We consider here the critical potential $V=\frac{g_*}{(2n)!} \phi^{2n}$ with $n$ integer for a family of multi-critical unitary theories, mostly following~\cite{Codello:2017qek}.

\noindent {\it - Field anomalous dimension $\gamma$}: 
it can be obtained by applying to the field two-point function 
$\langle \phi_x \phi_y\rangle = \frac{\tilde c}{|x-y|^{2\Delta}}$ ($\tilde c$ reduces to $c$ in the free theory)
the SDE twice,
\be \label{2}
\Box_x\Box_y \frac{\tilde c}{|x-y|^{2\Delta}}=\langle V'(\phi_x) V'(\phi_y) \rangle.
\ee
This gives at LO the value for $\gamma$:
\be\label{eq:anomalous-dimension}
\frac{16\delta_c(\delta_c+1)\gamma}{|x-y|^{2\delta_c+4}} c \;\stackrel{\mathrm{LO}}{=}\; 
\frac{1}{(2n-1)!} \frac{c^{2n-1}}{|x-y|^{2\delta_c+4}} \left( V_*^{(2n)} \right)^2 \,\, \Rightarrow
\boxed{\gamma = \frac{(n-1)^2}{8(2n)!}\,c^{2(n-1)} \left( V_*^{(2n)} \right)^2 . }
\ee

\noindent {\it - Anomalous dimension $\gamma_2$}:  for $n>2$ it can be derived applying twice the SDE to the three-point correlator $\langle\phi_x \,\phi_y\, \phi_z^{2}  \rangle$ (for $n=2$ SDE are applied once). Starting from
$
\Box_x\Box_y \langle \phi_x \,\phi_y\, \phi_z^{2}  \rangle=\langle V'(\phi_x) V'(\phi_y) \phi_z^{2} \rangle
$
one gets
\be
C^{\mathrm{free}}_{1,1,2} \,\frac{8(n-2)(\gamma_2 - 2\gamma)}{(n-1)^2}  \frac{1}{|x-y|^4|x-z|^{2\delta_c}|y-z|^{2\delta_c}} 
\,\stackrel{\mathrm{LO}}{=}\, \frac{ \left( V_*^{(2n)} \right)^2}{(2n-1)!^2} \frac{C^{\mathrm{free}}_{2n-1,2n-1,2}}{|x-y|^4|x-z|^{2\delta_c}|y-z|^{2\delta_c}}\,,
\ee
which defines $\gamma_2$.

\noindent {\it - Anomalous dimension $\gamma_k$}:  for $k\ge n$ it can be obtained applying once the SDE to the three-point correlator $\langle\phi_x \,\phi_y^k\, \phi_z^{k+1}\rangle$. The relation
$
\Box_x \langle \phi_x \,\phi_y^k\, \phi_z^{k\!+\!1}  \rangle=\langle V'(\phi_x) \,\phi_y^k\, \phi_z^{k\!+\!1} \rangle
$
gives at LO
\be
\frac{2}{n\!-\!1}(\gamma_{k+1}\!-\!\gamma_{k}\!-\!\gamma_1) \frac{C_{1,k,k+1}}{ |x-y|^{2} |y-z|^{\frac{2k}{n-1}-2} |z-x|^{\frac{2n}{n-1}}} \,\stackrel{\mathrm{LO}}{=}\,
\frac{V_*^{(2n)}  }{(2n\!-\!1)!} \frac{C_{2n-1,k,k+1}^{\mathrm{free}}}{ |x-y|^{2} |y-z|^{\frac{2k}{n-1}-2} |z-x|^{\frac{2n}{n-1}}}\,,
\ee
which results in the recurrence relation (with boundary condition $\gamma_{n\!-\!1}\stackrel{\mathrm{LO}}{=}0$)
\be
\gamma_{k+1} -\gamma_{k}= c_{n,k} V_*^{(2n)}+O\left[ \left( V_*^{(2n)}\right)^2\right]\
\, \Rightarrow
\boxed{\gamma_k = \frac{(n-1)c^{n-1}}{2 n!^2} \frac{k!}{(k\!-\!n)!} V_*^{(2n)}  }\, ,
\label{anomcompo}
\ee
where
$c_{n,k}=\frac{1}{ 2(n\!-\!2)! \, n!} \frac{k!}{(k\!-\!n\!+\!1)!} c^{n-1} $.
One must note that for $k=2n\!-\!1$ the SDE imply that $\Delta_{2n-1}=2+\Delta$, i.e. $\gamma_{2n-1}=\gamma+(n-1) \ep$ and that even if $V'(\phi)$ is a descendant operator,
at leading order three point correlators including this one satisfy the same CFT constraints as for a primary operator.

\noindent {\it - Structure constant $C_{1,2p,2q-1} $}:  it can be obtained for unitary theories with even interactions applying once the SDE to the three-point correlator $\langle\phi_x \,\phi_y^{2p}\, \phi_z^{2q-1}\rangle$. The relation
$
\Box_x \langle\phi_x \,\phi_y^{2p}\, \phi_z^{2q-1}\rangle=\langle V'(\phi_x) \,\phi_y^{2p}\, \phi_z^{2q-1}\rangle
$
gives at LO, removing the space time dependence, 
\be
 C_{1,2p,2q-1} (p\!-\!q)(p\!-\!q\!+\!1)(d_c\!-\!2)^2
  \stackrel{\mathrm{LO}}{=}
\frac{V_*^{(2n)} C^{\mathrm{free}}_{2n-1,2p,2q-1} }{(2n\!-\!1)!} 
\, \Rightarrow
\boxed{ C_{1,2p,2q-1}= \frac{V_*^{(2n)}(n\!-\!1)^2C^{\mathrm{free}}_{2n-1,2p,2q-1} }{4(p\!-\!q)(p\!-\!q\!+\!1)(2n\!-\!1)!}  \,,}
\nn
\ee
which is valid in the range $q+p \geq n$, $q-p \geq 1-n$ and $q-p \neq 0,1$. 

Similar results for $C_{1,2p,2q} $ and $C_{1,2p-,2q-1} $ for non unitary odd models can be obtained following the same procedure.

\noindent {\it - Structure constant  $C_{1,1,2k} $}:  it can be obtained applying twice the SDE to the three-point correlator $\langle\phi_x \,\phi_y\, \phi_z^{2k}\rangle$. The relation
$
\Box_x \Box_y \langle\phi_x \,\phi_y\, \phi_z^{2k}\rangle=\langle V'(\phi_x) \,V'(\phi_y)\, \phi_z^{2k}\rangle
$
becomes at LO, removing the space time dependence, 
\be
\frac{16k(k\!-\!1)(k\!-\!n)(k\!-\!n\!+\!1)}{(n\!-\!1)^4} C_{1,1,2k}
 \stackrel{\mathrm{LO}}{=}
 \frac{ \left(V_*^{(2n)}\right)^2 }{(2n\!-\!1)!^2} C_{2n-1,2n-1,2k}^{\mathrm{free}}\,,
\nn
\ee
from which one can derive the expression for $C_{1,1,2k}$, valid for $2\leq k\le 2n-1$ and $k\neq n-1,n$.

\noindent {\it - Criticality condition}:  It can be obtained in different  equivalent ways. Let us derive it analyzing the correlator
$\langle\phi_x \,\phi_y\, \phi_z^{2n-2}\rangle$, for which the structure constant $C_{1,1,2n-2}$ can be obtained from
$C_{1,2p,2q-1}$ setting $q=1$ and $p=n-1$
\be
C_{1,1,2n-2}   \stackrel{\mathrm{LO}}{=} \frac{(n-1)c^{2n-1}}{4(n-2)} V_*^{(2n)}.
\ee
Starting from the relation
$
\Box_x \Box_y \langle\phi_x \,\phi_y\, \phi_z^{2n-2}\rangle=\langle V'(\phi_x) \,V'(\phi_y)\, \phi_z^{2n-2}\rangle
$
one obtains
\be
\frac{8(n\!-\!2)}{(n\!-\!1)^2} \left((n\!-\!1)\epsilon\!-\!\gamma_{2n-2}\right)\frac{C_{1,1,2n-2}}{|x-y|^{2\delta_c+2}|y-z|^2|z-x|^2} 
  \stackrel{\mathrm{LO}}{=}
\frac{\left(V_*^{(2n)}\right)^2}{(2n-1)!^2}  \frac{C^{\mathrm{free}}_{2n-1,2n-1,2n-2}}{|x-y|^{2\delta_c+2}|y-z|^2|z-x|^2},
\ee
which gives
\be
\frac{2}{n-1}\left((n-1)\epsilon-\gamma_{2n-2}\right) V_*^{(2n)}
  \stackrel{\mathrm{LO}}{=}
 \frac{(2n-2)!c^{n-1}}{n!(n-1)!^2}\left( V_*^{(2n)} \right)^2
 \ee
 and substituting the LO expression for the anomalous dimension $\gamma_{2n-2}$,
 \be\label{eq:fp-general}
 0 = (1-n)\epsilon\, V_*^{(2n)} + \frac{(n-1)(2n)!}{4n!^3}c^{n-1} \left( V_*^{(2n)} \right)^2\,,
\ee
which fixes the dependence in $\ep$ of the critical coupling $g_*=V_*^{(2n)}$.

\subsection{Relation to the FPRG approach}
We consider here the unitary multi-critical single field scalar theories to LO with marginal interaction $\varphi^{2n}$. 
Results obtained in the CFT-SDE approach for the criticality condition and for the anomalous dimensions of the composite operators, 
which can be written as monomial $\phi^k$ for $k\ge n$ to LO,
can be conveniently combined together in a single functional relation which is the same obtained in a perturbative RG approach raised at functional level, the FPRG flow equation. The construction goes as follows. 
First of all it is convenient to redefine the quantities rescaling $V\rightarrow 4V/\left( (n-1)c^{n-1}\right)$.

The criticality condition, after rescaling the coupling $g_*=V_*^{(2n)}=v_*^{(2n)}$, reads
\be
0 = (1-n)\epsilon\, v_*^{(2n)} + \frac{(2n)!}{n!^3} \left(v_*^{(2n)}\right)^2.
\ee
Diving by $(2n)!$ and multiplying by $\varphi^{2n}$, where the rescaled field $\varphi=\mu^{-\delta}\phi$, one gets a more suggestive form
\be
0 = (1-n)\epsilon\, \frac {v_*^{(2n)}}{(2n)!}\varphi^{2n} + \frac{1}{n!} \left(\frac{v_*^{(2n)}}{n!} \varphi^{n} \right)^2 ,
\ee
so that, on defining the critical multicritical potential as
$v_*(\varphi)=\frac {v_*^{(2n)}}{(2n)!}\varphi^{2n}$, such that $v_*(\varphi)=\mu^{-d} V_*(\mu^\delta \varphi)$, one can write the criticality condition as
\be
0 = -d\, v_* +\left(\frac{d}{2}-1\right) \varphi \,v_*' + \frac{1}{n!} \left(v_*^{(n)} \right)^2\,,
\label{FP_FPRG}
\ee
since $d=\frac{2n}{n-1}-\epsilon$.
This is the LO fixed point equation obtained in a perturbative framework in the $\overline{\rm MS}$ scheme where the coupling (critical potential) has been conveniently rescaled.
Indeed the term proportional to the anomalous dimension of the field $\gamma \sim O(\epsilon^2)$, 
which would be given by $\gamma \, \varphi \, v_*'$, is negligible at this order, since $v_*^{(2n)}\sim O(\epsilon)$.

We can now move on and consider the additional information given by the anomalous dimension of the composite operators $[\phi^i]$ obtained in Eq.~\eqref{anomcompo} to which we apply the rescaling mentioned above,
\be
\gamma_i= \frac{2}{n!^2} \frac{i!}{(i-n)!} v_*^{(2n)}\quad , \quad i\ge n
\label{anomdim_i}
\ee
Therefore the operator $[\phi^i]$ has a scaling dimension $\Delta_i= \left(\frac{d}{2}-1\right)  i +\gamma_i$. 
To parametrize a deformation around the multi-critical theory along these directions one can introduce the corresponding couplings $g_i$,
which have dimensions $\theta_i=d-\Delta_i$.
This means that the linearized flow around the fixed point induced by a scale change, must be
\be \label{eq:linearized-flow}
\mu \frac{d}{d \mu} g_i=-\theta_i g_i = \left(-d +\left(\frac{d}{2}-1\right)  i +\gamma_i\right) g_i.
\ee
Introducing the quantity $\delta v_i=\frac{g_i}{i!} \varphi^i$ and substituting the value of the LO anomalous dimension
$\gamma_i$ from Eq.~\eqref{anomdim_i}, one can write

\be
\mu \frac{d}{d \mu} \delta v_i= -d \delta v_i +\left(\frac{d}{2}-1\right)  \varphi \delta v_i' + \frac{2}{n!^2} \frac{i!}{(i-n)!} v_*^{(2n)} \delta v_i
\ee
and noting that the last term can be rewritten as
\be
\frac{2}{n!^2} \frac{i!}{(i-n)!} v_*^{(2n)} \frac{g_i}{i!} \varphi^i= \frac{2}{n!} \frac{ v_*^{(2n)}}{n!} \varphi^n \frac{g_i}{(i-n)!} \varphi^{i-n}= \frac{2}{n!} v_*^{(n)} \delta v_i^{(n)}
\simeq \frac{1}{n!} \left[  \left( v_*^{(n)}+\delta v_i^{(n)}\right)^2 - (v_*^{(n)})^2 \right], 
\ee
where $i\!-\!n\ge 0$, one can pack the information of the critical condition and of the flow for any power-like deformation at LO in a single equation.
Indeed defining the potential $v=v_*+ \sum_i \delta v_i$ and taking into account Eq.~\eqref{FP_FPRG} one can write
\be \label{eq:beta-functional}
\mu \frac{d}{d \mu} v=-d \, v +\left(\frac{d}{2}-1\right)  \varphi \, v'+\frac{1}{n!} \left(v^{(n)}\right)^2\,.
\ee
This is the so called functional perturbative RG flow equation for the potential~\cite{Codello:2017hhh, ODwyer:2007brp}, 
restricted at LO so that it takes into account only the $O(\epsilon)$ corrections.
It is interesting to note that such parallelism among CFT and perturbation theory is still valid at NLO where the field anomalous dimension start to play an important role. 
In this case additional information is also given by some special structure constants (OPE coefficients) derived at LO in the CFT+SDE framework
which can be also obtained analyzing the expansion of the beta functional for $v$ in the second order in the deformations~\cite{Codello:2017hhh}. 
We also note that for non unitary theories, with standard kinetic terms but odd potential interactions or higher derivative theories which can have also derivative interactions,
results in the $\ep$-expansion obtained assuming conformal symmetry have been found to be in full agreement with renormalization group analysis~\cite{Safari:2017irw, Safari:2017tgs}.
 
\subsection{Example: the universality class of the critical Ising model in $d<4$} \label{sect:example1}

Here we specialize some of the results of this section to a critical field theory with $\varphi^4$ interaction in $d<4$ dimensions.
Since the model is known to capture the physics of the universality class of the lattice Ising model at criticality, we will make explicit connection with the language of statistical
field theory. We take the potential to be $v=\frac{\lambda}{4!}\varphi^4$, therefore using $d=4-\epsilon$ and $n=2$
in \eqref{eq:beta-functional}, which means that a rescaling $v\to \frac{4}{c} v$ is understood, we find that the flow of the coupling $\lambda$ becomes
the well-known beta function
\be \label{eq:betarescaledising}
 \beta_\lambda = -\epsilon \lambda + 3\lambda^2\,.
\ee
Using \eqref{eq:anomalous-dimension} and \eqref{eq:fp-general} (or ~\eqref{eq:anomalous-dimension} with a rescaling $v\to \frac{4}{c} v$ together with~ \eqref{eq:betarescaledising})  
we find the anomalous dimension of the field $\gamma$ which is related to the critical exponent $\eta=2\gamma$.
Explicitly
\be 
 \eta = \frac{\epsilon^2}{54}\,,
\ee
which is quadratic in $\epsilon$ as expected, and therefore gives a subleading contribution to the problem.
The scaling exponent $\gamma_2$ which corrects the scaling dimension $\Delta_2$ of the composite quadratic operator $[\phi^2]$ is obtained using \eqref{anomdim_i} for $i=2$.
By definition $\gamma_2=\lambda=\ep/3$ is related to the critical exponent $\theta_2=2-\frac{\epsilon}{3}$ governing the scaling of the correlation length $\nu = \theta_2^{-1}$,
which can be determined using \eqref{eq:linearized-flow} and some standard hyperscaling arguments.
We find to the leading order
\be 
 \nu = \frac{1}{2}+\frac{\epsilon}{12}\,,
\ee
which completes the determination of the independent (infrared relevant) critical exponents governing the critical point of the Ising universality class.
All subsequent thermodynamical exponents can be deduced using the hyperscaling hypothesis.

\section{Multiple scalar field case}

The analysis for theories with multiple scalar fields goes along the lines of the one briefly recalled in the previous Section,
but with some important differences which requires some extesions~\cite{Codello:2018nbe}. One can summarize them
\begin{itemize}
\item The scaling scalar fields arise from a mixing which is manifest in general in the splitting induced by different anomalous dimensions $\gamma_i$, eigenvalues of the anomalous dimension matrix.
\item There are  many more composite operators (even without derivatives) at LO, defined in general as a superposition of all 
monomials of the scaling fields $\cS_{k}=S_{i_1 \cdots i_k} \phi_{i_1} \cdots \phi_{i_k}$ with scaling dimension 
$\Delta^S_k= k \,\delta+\gamma^S_k$. 
The analysis leads to recurrence relations which can be solved to give secular equations for the LO anomalous dimensions 
$\gamma^S_k$ (eigenvalues) and the tensors $S_{i_1 \cdots i_k}$ (eigenvectors).
\item Structure constants (involving some composite operators  $\cS_{k}$) are obtained just as in the single field case.
\item For unitary models with even interactions and models with cubic interactions~\footnote{In the multi-field case critical models with cubic interactions can be either unitary or non unitary in perturbation theory.} ($d_c=6$) one can obtain as in the single field case the criticality conditions and see that the relation with the FPRG approach.
\end{itemize}
In the following the main results are illustrated. We refer to~\cite{Codello:2018nbe} for the details of the derivations.
\subsection{Field anomalous dimensions}
For the primary fields $\phi_i$ the two-point function is constrained by CFT as
\be 
\langle \phi_i(x) \phi_j(y)\rangle = \frac{\tilde c\delta_{ij}}{|x-y|^{2\Delta_i}} ,
\label{2Pc}
\ee
where $\tilde c$ is a constant whose value in the free theory is $c$.
Applying the SDE one finds that the anomalous dimensions are given by the eigenvalues of the matrix
\be\label{eq:anomalous-dimension-multifield}
\gamma_{ab} = \frac{(n-1)^2}{8(2n)!}\,c^{2(n-1)} V_{ai_1i_2\cdots i_{2n-1}} V_{bi_1i_2\cdots i_{2n-1}},
\ee
which is valid for  both integer (unitary theories)  and semi-odd (perturbatively unitary or non unitary theories) values of $n$.
The matrix depends on the particular solution for the critical theory in theory space one considers. Depending on the level of symmetry of the critical theory (the critical potential) anomalous dimensions can be all different or have various levels of degeneracy.
\subsection{Anomalous dimensions for composite operators}

\noindent {\it - Quadratic operators}: they can be studied exploiting the properties of the three-point correlator

$\left\langle \phi_i(x) \phi_j(y) [S_{pq} \phi_p\phi_q](z) \right\rangle$.
For $n=2$ consistency by applying the SDE once gives the secular equation
\be
\gamma^S_2  \, S_{ij} = \frac{c}{4} V_{ijab}\,S_{ab}.
\ee
In the other cases one has to apply the SDE twice and obtains
\bea
\gamma^S_2\, S_{ij} = \frac{(n\!-\!1)^2c^{2(n-1)}}{8(n\!-\!2)(2n\!-\!2)!}\, V_{i\,p\,i_2\cdots i_{2n-1}}V_{j\,q\,i_2\cdots i_{2n-1}} \,S_{pq} 
+ \frac{(n\!-\!1)^2c^{2(n-1)}}{8(2n)!}\, \left( V_{i\,i_1\cdots i_{2n-1}}V_{p\,i_1\cdots i_{2n-1}} \,S_{pj} + i \leftrightarrow j \right). \nn
\label{gammaS2}
\eea
Moreover of $n=3/2$ there is a family of descendant scaling operator given by $V_i(\phi)=\frac{d}{d \phi_i} V$ with anomalous dimension $\gamma_2^i=\gamma_i+\ep/2$.

\noindent {\it - Higher order operators}:  the LO anomalous dimensions for the scaling operators $\cS_{k}$ with $k \ge n$
can be extracted from the study of the three-point correlators $\langle \phi_i(x) \, \cS_{k}(y) \cS_{k+1}(z) \rangle$ by applying once the SDE and impose the consistency. The relation
$
\Box_x \langle \phi_i(x) \, \cS_{k}(y) \cS_{k+1}(z) \rangle=\langle V_i(\phi(x) ) \, \cS_{k}(y) \cS_{k+1}(z) \rangle
$
gives at LO
\be
 \frac{(\gamma^S_{k+1}\!-\!\gamma^S_k\!-\!\gamma_i)}{n\!-\!1} \frac{2 C^{\mathrm{free}}_{1,k,k+1} \, S_{ii_1\cdots i_k}S_{i_1\cdots i_k}}{|x-y|^2|y-z|^{2k\delta_c-2}|z-x|^{2\delta_c+2}}
\stackrel{\mathrm{LO}}{=} \frac{C^{\mathrm{free}}_{2n-1,k,k+1}}{(2n-1)!} \,  \frac{V_{ii_1\cdots i_rj_1\cdots j_s}S_{j_1\cdots j_s l_1\cdots l_t}S_{l_1\cdots l_ti_1\cdots i_r}}{|x-y|^2|y-z|^{2k\delta_c-2}|z-x|^{2\delta_c+2}}\,,
\ee
leading to the recurrrence relation
\be \label{rr} 
(\gamma^S_{k+1}-\gamma^S_k-\gamma_i) S_{ii_1\cdots i_k}S_{i_1\cdots i_k} = c_{n,k} V_{ii_1\cdots i_rj_1\cdots j_s}S_{j_1\cdots j_s l_1\cdots l_t}S_{l_1\cdots l_ti_1\cdots i_r}\,,
\ee
where $c_{n,k}$ has been defined in the previous Section.
To solve this recurrence relation one has to find before $\gamma^S_n$ and then proceed by induction. In doing so some algebraic manipulations are necessary in order to be able to single out, from the secular equation for $\gamma^S_k$ and $S_{i_1\cdots i_k}$, a contraction with the tensor $S_{i_1\cdots i_k}$ itself, so that the equation becomes linear in the tensor $S$. Details can be found in~\cite{Codello:2018nbe}.
Observing that $\gamma^S_k$, for $k<n$ are subleading (depend quadratically in the critical potential) one can find
\be \label{gamma-S-n} 
\gamma^S_n S_{i_1\cdots i_n} = c_{n,n-1} V_{i_1\cdots i_nj_1\cdots j_n}S_{j_1\cdots j_n}
\ee
and for the full infinite tower of this family of composite operators ($l\ge0$)
\be  
\gamma^S_{n+l} S_{i_1\cdots i_{n+l}} = \frac{(n-1)c^{n-1}}{2n!^2} \frac{(n+l)!}{l!}\, V_{j_1\cdots j_n(i_1\cdots i_n}S_{i_{n+1}\cdots i_{n+l})j_1\cdots j_n} \label{gamma-k}\,,
\ee
where the round backets stand for the symmetrization of the enclosed indices.
One can verify that the LO recurrence relation is valid also for $k=2n-2, 2n-1$, cases for which in the three-point correlator a descendant operator is present, fact which makes impossible to use the form of Eq.~\eqref{cft-3pf}. 
This problem can be bypassed using again the SDE and this suggests also a way to find for the unitary theories the criticality condition, which gives the solutions for critical potential (couplings) as functions of $\ep$.

\subsection{Structure constants}
The computation of structure constants for the multi-field case is a straightforward extension of the single-field case
described in section 3.

\noindent {\it - Structure constants $C_{\phi_i \cS_{2p} \tilde{\cS}_{2q-1}}$ }:  for unitary theories with even interactions they can be obtained applying once the SDE to the three-point correlator $\left\langle \phi_i(x) \cS_{2p}(y) \tilde{\cS}_{2q-1}(z) \right\rangle $. Indeed the relation

$
\Box_x \left\langle \phi_i(x) \cS_{2p}(y) \tilde{\cS}_{2q-1}(z) \right\rangle=
\left\langle V_i(\phi(x)) \cS_{2p}(y) \tilde{\cS}_{2q-1}(z) \right\rangle
$
gives at LO
\be  \label{ciuv-even}
C_{\phi_i \cS_{2p} \tilde{\cS}_{2q-1}} = \frac{V_{il_1\cdots l_r k_1\cdots k_t}\,S_{j_1\cdots j_s l_1 \cdots l_r} \, \tilde{S}_{k_1\cdots k_t j_1 \cdots j_s}}{(2n-1)!} \frac{(n-1)^2}{4(p-q)(p-q+1)}C^{\mathrm{free}}_{2n-1,2p,2q-1}\,,
\ee  
which is valid in the range $q+p \geq n$, $q-p \geq 1-n$ and $q-p \neq 0,1$, with $2n\!-\!1= r\!+\!t$, $2p = r\!+\!s$ and $2q\!-\!1 = s\!+\!t$.
The $S$ and $\tilde{S}$ tensors are among the solutions of the secular equations of the previous Section.

\noindent {\it - Structure constants $C_{\phi_i \cS_{2p} \tilde{\cS}_{2q}}$ and $C_{\phi_i \cS_{2p-1} \tilde{\cS}_{2q-1}}$}: they can be obtained for theories with odd interactions in a similar way. Setting $l=n-1/2$:
\be  \label{ci2p2q}
C_{\phi_i \cS_{2p} \tilde{\cS}_{2q}}  = \frac{V_{il_1\cdots l_r k_1\cdots k_t}\,S_{j_1\cdots j_s l_1 \cdots l_r} \, \tilde{S}_{k_1\cdots k_t j_1 \cdots j_s}}{(2\ell)!} \frac{(2\ell-1)^2}{4(4(p-q)^2-1)}C^{\mathrm{free}}_{2\ell,2p,2q},
\ee 
which is valid only in the range $q+p\geq \ell$ and $|q-p|\leq \ell$ and for $2\ell= r+t$, $2p = r+s$, $2q = s+t$.
Making the shift $p\rightarrow p-\frac{1}{2}$ and $q\rightarrow q-\frac{1}{2}$ one finds also
\be  \label{ci2p-12q-1}
C_{\phi_i \cS_{2p-1} \tilde{\cS}_{2q-1}} = \frac{V_{il_1\cdots l_r k_1\cdots k_t}\,S_{j_1\cdots j_s l_1 \cdots l_r} \, \tilde{S}_{k_1\cdots k_t j_1 \cdots j_s}}{(2\ell)!} \frac{(2\ell-1)^2}{4(4(p-q)^2-1)}C^{\mathrm{free}}_{2\ell,2p-1,2q-1},
\ee 
where now $q,p$ fall in the range $q+p\geq \ell+1$ and $|q-p|\leq \ell$, and the integers $r,s,t$ satisfy the relations
$2\ell= r+t$, $2p-1 = r+s$, $2q-1 = s+t$.

\noindent {\it - Structure constants  $C_{\phi_i,\phi_j,\cS_{2p}} $}:  they can be obtained applying twice the SDE to the three-point correlator $\langle\phi_i(x) \,\phi_j(y)\, \phi_z^{2k}\rangle$. Evaluating
$
\Box_x \Box_y \langle\phi_i(x) \,\phi_j(y)\, \cS_{2k}(z) \rangle=\langle V_i(\phi(x)) \,V_j(\phi(y))\, \cS_{2k}(z) \rangle
$
one obtains
\bea
C_{\phi_i \phi_j \cS_{2k}} = \frac{(n-1)^4c^{2n+k-1}}{16k(k-1)(k-n)(k-n+1)} \frac{(2k)!}{k!^2(2n-k-1)!^2} 
V_{ii_1\cdots i_{2n-k-1}a_1\cdots a_k}V_{ji_1\cdots i_{2n-k-1}b_1\cdots b_k}S_{a_1\cdots a_k \,b_1\cdots b_k},\nn
\label{ciju}
\eea
valid for $2\leq k\le 2n-1$ and $k\neq n-1,n$.

\noindent {\it - Structure constants  $C_{\phi_i,\phi_j,\phi_k}$}:  for theories with odd interactions, i.e. with $m=2n$ odd (we define $n=l\!+\!1/2$), they can be obtained applying three times the SDE to the three-point correlator $\langle\phi_i(x) \,\phi_j(y)\, \phi_k(z)\rangle$.
The relation
$
\Box_x \Box_y \Box_z \langle\phi_i(x) \,\phi_j(y)\, \phi_k(z)\rangle=\langle V_i(\phi(x)) \,V_j(\phi(y))\, V_k(\phi(z))\rangle
$
becomes at LO
\bea
\frac{2^8\ell(\ell-1)}{(2\ell-1)^6} \frac{C_{\phi_i \phi_j \phi_k}}{|x-y|^{\delta_c+2}|x-z|^{\delta_c+2}|y-z|^{\delta_c+2}}
\stackrel{\mathrm{LO}}{=}\frac{V_{i\,a_1\cdots a_\ell\,b_1\cdots b_\ell}V_{j\,b_1\cdots b_\ell\,c_1\cdots c_\ell}V_{k\,c_1\cdots c_\ell\,a_1\cdots a_\ell}}{(2\ell)!^3 |x\!-\!y|^{2\ell\delta_c}|x\!-\!z|^{2\ell\delta_c}|y\!-\!z|^{2\ell\delta_c}}
C^\mathrm{free}_{2\ell,\,2\ell,\,2\ell}.\nn
\eea
Noting that $2\ell\delta_c = \delta_c + 2$ one obtains
\be \label{cijk}
C_{\phi_i \phi_j \phi_k} = \frac{(2\ell-1)^6c^{3\ell}}{2^8\ell(\ell-1)\ell!^3}\,V_{i\,a_1\cdots a_\ell\,b_1\cdots b_\ell}V_{j\,b_1\cdots b_\ell\,c_1\cdots c_\ell}V_{k\,c_1\cdots c_\ell\,a_1\cdots a_\ell}.
\ee

\subsection{Criticality conditions}
\noindent {\it - $d_c=6$  or $n=3/2$}:  this case corresponds to theories with cubic interactions. 

\noindent The relation
$
\Box_x \Box_y \Box_z \langle\phi_i(x) \,\phi_j(y)\, \phi_k(z)\rangle=\langle V_i(\phi(x)) \,V_j(\phi(y))\, V_k(\phi(z))\rangle
$
gives at LO
\be
 \frac{32(\epsilon\!-\!2(\gamma_i\!+\!\gamma_j\!+\!\gamma_k))}{|x\!-\!y|^4|y\!-\!z|^4|x\!-\!z|^4}C_{\phi_i\phi_j\phi_k} 
 \stackrel{\mathrm{LO}}{=}
 \frac{c^3\, V_{iab} V_{jbc}V_{kca}}{|x-y|^4|y-z|^4|x-z|^4}\,,
\ee
where $C_{\phi_i\phi_j\phi_k} = -\frac{c^2}{4} V_{ijk}$ can be obtained from Eq.~\eqref{ci2p-12q-1} for $l=p=q=1$.
Then one finds the equation
\be 
\boxed{
8(2(\gamma_i+\gamma_j+\gamma_k)-\epsilon) V_{ijk} = c\, V_{iab} V_{jbc} V_{kca}.
}
\ee
This condition can be obtained from the Functional perturbative RG equation at LO for the potential
\be 
\beta_v = -d v +\frac{d-2}{2}\phi_i v_i + \phi_i \gamma_{ij} v_j-\frac{2}{3} v_{ij} v_{jl} v_{li}\,.
\ee
adopting the diagonal basis where $\gamma_{ij}\rightarrow \gamma_i\delta_{ij}$.

\noindent {\it - $d_c=4$  or $n=2$}: this case corresponds to unitary theories with quartic interactions. The relation can be found applying twice the SDE to the
three point correlator $\langle\phi_i(x) \,\phi_j(y)\, \cS_2(z)\rangle$ and making use of the secular equation for the quadratic primary composite operators (at LO)
$\gamma^S_2  \, S_{ij} = \frac{c}{4} V_{ijab}\,S_{ab}$, which can be obtained setting $n=2$ and $l=0$ in Eq.~\eqref{gamma-k}.
From the SDE equations one obtains
\be 
8c^2\gamma^S_2(\epsilon-\gamma^S_2) S_{ij}=c^4 V_{pqik} V_{pqjl} S_{kl}\,
\ee
and using twice the secular relation involving $\gamma^S_2$ this relation becomes
\be 
\boxed{
-\epsilon\,\, V_{ijkl}  +\frac{c}{4} \left(V_{pqik}V_{ljpq} + V_{pqil}V_{kjpq} + V_{ijpq} V_{pqkl}\right)  =0\,,}
\ee
having factored out the dependence in the symmetric tensors, which span the whole space of symmetric objects with two indices.

\noindent {\it - Integer $n$}:  this case corresponds to unitary multi-critical theories with even interactions. The criticality condition can be obtained
applying twice the SDE to the three point correlator $\langle\phi_i(x) \,\phi_j(y)\, \cS_{2n-2}(z)\rangle$ and making use of the secular equation given in Eq.~\eqref{gamma-k}.
Evaluating the relation
$
\Box_x \Box_y \langle \phi_i(x) \,\phi_j(y)\, \cS_{2n-2}(z) \rangle=\langle V_i(\phi(x)) \,V_j(\phi(y))\, \cS_{2n-2}(z) \rangle
$
one finds
\be
\frac{2}{n\!-\!1} \left((n\!-\!1)\epsilon\!-\!\gamma_{2n-2}^S\right)  V_{ijl_1\cdots l_{2n-2}} S_{l_1\cdots l_{2n-2}} 
\stackrel{\mathrm{LO}}{=}
 \frac{(2n-2)!c^{n-1}}{n!(n-1)!^2} V_{ii_1\cdots i_{n-1}j_1\cdots j_n} V_{j j_1\cdots j_n l_1\cdots l_{n-1}} S_{l_1\cdots l_{n-1}i_1\cdots i_{n-1}}\,.\nn
\ee
Then one uses the eigenvalue equation in Eq.~\eqref{gamma-k} for $l=n\!-\!2$ and after few algebraic manipulation where we also factorize the dependence in the symmetric tensor $S_{i_1\cdots i_{2n-2}}$ one finally obtains
\be \label{fp-even}
\boxed{
0 = (1-n)\epsilon\, V_{i_1\cdots i_{2n}} + \frac{(n-1)(2n)!}{4n!^3}c^{n-1} V_{j_1\cdots j_n(i_1\cdots i_n} V_{i_{n+1}\cdots i_{2n})j_1\cdots j_n}.} 
\ee
Similarly to the single field case one can show that the criticality condition as well as the secular equations for the composite operators given above can be obtained from the Functional Perturbative RG equation at LO for the potential~\cite{Osborn:2017ucf, Zinati:2019gct}
\be 
 \label{eq:beta-functional-multifield}
\beta_v = -d v +\frac{d-2}{2}\phi_i v_i +\frac{1}{n!} v_{j_i\cdots j_n} v_{j_i\cdots j_n}\,.
\ee

\subsection{Example: the universality class of the critical $O(2)$ Heisenberg model in $d<4$}

Here we briefly discuss a simple application of the results of this section to the critical $O(2)$ Heisenberg model.
Like in the example of section \ref{sect:example1} we specialize to $n=2$ and therefore to a quartic interaction. We also choose a
total of $N=2$ fields $\phi=\left(\phi_1,\phi_2\right)$. The maximal symmetry that the model can have is $O(2)$,
which is the symmetry content of the Heisenberg model at criticality.
The potential is constrained to be of the invariant form $v=\frac{\lambda}{4!}\left(\varphi_1^2+\varphi_2^2\right)^2$
because it depends on the $O(2)$ invariant order parameter $\rho=\varphi_1^2+\varphi_2^2$.
Using \eqref{eq:beta-functional-multifield} the beta function is
\be \label{eq:betaheisemberg}
 \beta_\lambda = -\epsilon \lambda + \frac{10}{3}\lambda^2\,.
\ee
Using \eqref{eq:anomalous-dimension-multifield}, where se should take into account the rescaling $v\to \frac{4}{c} v$ also used for~\eqref{eq:betaheisemberg}, and expressing the derivatives of the potential we can determine the
anomalous dimension matrix $\gamma_{ab}$, which in general must be diagonalized.
However in this case $\gamma_{ab}$ is a two-by-two diagonal matrix, so we can evince the anomalous dimension $\eta$
directly from $\eta \delta_{ab}=2\gamma_{ab}$. 

We find
\be 
 \eta = \frac{\epsilon^2}{50}\,,
\ee
which is obviously shared by all fields.
The spectrum of composite operators containing two copies of the fields is more complicate sice it contains operators that violate
the model's symmetry. One operator is however invariant and, in fact, coincides with the order parameter.
Using the same logic of section \ref{sect:example1} from the scaling of this operator we can determine the scaling exponent of the correlation length
\be 
 \nu = \frac{1}{2}+\frac{\epsilon}{10}\,.
\ee
Interestingly our very general approach gives immediate access to all the deformations of the model which are not $O(2)$ invariant
and which might, consequently, be forgotten in approaches that make more use of symmetry constraints. As to prove this point
we briefly mention that the above analysis can be generalied to $N>2$ arbitrary components, and that the corresponding field theory can have
both a Heisenberg-type critical point which is maximally $O(N)$-invariant and a ``cubic'' anisotropic point which is not. In this case it is crucial to understand the role
of the symmetry breaking deformations of the potential, in order to answer the question on which of the two is the IR critical point
of the theory \cite{Kleinert:2001ax}.

%

\section{Discussion}
As it has been recalled above, several universal data of the critical theories can be obtained in a perturbative approach at the first non trivial order by assuming the 
theory to be conformal and making use of the SDE which follow from an available lagrangian description at criticality (as in the Landau-Ginzburg approach) and
in particular this is true also for multi-field theories.
In almost all investigations present in the literature, either using CFT or RG methods, perturbative or non perturbative, global symmetries in multi-field theories are assumed from the start, since this greatly constrains the number of possible interactions and therefore of possible critical theories. Such critical theories, not considering effects of symmetry breaking, have always a higher or at least equal symmetry compared to the assumed one. E.g. if one consider all possible scalar theories with symmetries $O(N)\times O(M)$ a fixed point with symmetry $O(N+M)$ exists. Similarly for theories with scalar and fermions critical theories with an (enhanced) supersymmetric sector can appear~\cite{Vacca:2015nta, Gies:2017tod}. For unitary models of this kind critical theories with enhanced symmetry are typically infrared attractive, meaning that these larger symmetries can be interpreted as an emergent phenomena at large distances.

In some past RG investigations of theories with quartic interactions the so called trace property condition for the critical potential,
which leads to full degeneracy of the field anomalous dimensions, was assumed therefore reducing the possible number of critical
theories~\cite{Wallace:1975ez, Brezin:1973jt, Michel:1983in}.
A general study without assumptions has yet to come.
An attempt to systematize this search for all possible theories with two fields has been done in~\cite{Osborn:2017ucf}.
Also studies of theories with cubic interactions can be interesting. Multi field theories with $N$ fields and $O(N-1)$ symmetry have been studied at large N in~\cite{Fei:2014xta} showing the appearance of a unitary critical theory at perturbative level. For $N=2$ this is the case also for the 3-state Potts model.
We have analyzed them up to $N=3$ and found six non trivial novel critical theories, with three real different field anomalous dimensions, or two degenerate or all degenerate. In particular two critical theories, characterized by some specific symmetries, appear to have all positive anomalous dimensions and therefore unitary at perturbative level. Similarly we have analyzed theories with quintic interaction and $N=2$. In our approach we can also show that there are no other theories with quartic interactions besides the ones already known in the literature. All these results will appear soon in a forthcoming work~\cite{CSVZ7}. We find also non unitary critical theories with not only complex couplings  but complex anomalous dimensions.
These are related to the idea of complex conformal field theories and can be relevant to describe properties of RG flows~\cite{Gorbenko:2018ncu, Gorbenko:2018dtm}. 

Even if certainly a difficult task, it is important to start systematic analysis of theories with several fields not assuming any symmetry and study the full spectrum of possible critical theories characterizing the theory space. In particular we have shown that the knowledge of the scaling dimension of composite operators can give access within a certain approximation, such as in perturbation theory in $\ep$-expansion, to RG flows properties inside the theory space.
We note that a flow trajectory which for any reason pass close to some critical point in theory space, characterized by a certain symmetry, is related to the fact that such a theory spends a large amount of RG-time inside a quasi conformal windows and is characterized by the corresponding approximate global symmetry.
One can envisage some interesting cases, which could be of interest in the quest of searching UV completion for the Standard Model (SM) of particle physics.
Independently of the UV completed model, which could be related also to an asymptotic safety scenario even not considering gravity as recently suggested~\cite{Litim:2014uca},
having a QFT description at some high but under Planck scale in some theory space, the RG flow could pass at some scale $M$ much larger than the Electroweak one ($M_{EW}<M<M_{Pl}$) close to fixed points which could be characterized by certain level of SUSY or even some GUT symmetry.
From lower energy scales, having at our disposal experimental scattering data with increasing energies, one could have the impression that the fundamental theory is characterized by such symmetries, even if they were just approximate in a certain energy window.
Moreover a renormalizable scenario which fits the asymptotic safety paradigm is related to a fixed point with a finite number of relevant operators (directions) in the theory space, so that it provides dynamically a high degree of predictivity since most of the parameters (couplings) of the theory are not independent along the flow. 
This shows the importance could have a systematic understanding of the theory space of the SM QFT, which we believe to be just an effective theory, or of some of its extensions, in building a comprehensive picture. All possible tools to characterize all the non trivial critical theories in four (and also other) dimension would be welcome.


\acknowledgments{G.~P.~V. , A.~C. and O.~Z. are grateful to Martin Reuter and all other partecipants of the workshop {\it Quantum Fields - From Fundamental Concepts to Phenomenological Questions} for the interest shown in the topics of this work. The research of O.~Z. was funded by Deutsche Forschungsgemeinschaft (DFG) under the Grant Za 958/2-1.}



%

\end{document}